\documentclass[]{spie}

\usepackage[]{graphicx,color}
\usepackage[]{amsmath}
\usepackage[]{amssymb}

\title{MEMS practice, from the lab to the telescope}

\author{
Katie M. Morzinski\authorinfo{\supit{*}Sagan Fellow.   Currently University of Arizona; formerly UC--Santa Cruz.  Contact: ktmorz@arizona.edu}\supit{*a,b},
Andrew P. Norton\supit{a,c},
Julia Wilhelmson Evans\supit{a},
Layra Reza\supit{a},
Scott A. Severson\supit{a,d},
Daren Dillon\supit{a,c},
Marc Reinig\supit{a,c},
Donald T. Gavel\supit{a,c},
Steven Cornelissen\supit{a,e},
Bruce A. Macintosh\supit{a,f},
and
Claire E. Max\supit{a,c}
\skiplinehalf
\supit{a} Center for Adaptive Optics, University of California, Santa Cruz, CA;
\skiplinehalf
\supit{b} Steward Observatory, University of Arizona, Tucson, AZ;
\skiplinehalf
\supit{c} Laboratory for Adaptive Optics, University of California, Santa Cruz, CA;
\skiplinehalf
\supit{d} Department of Physics and Astronomy, Sonoma State University, Rohnert Park, CA;
\skiplinehalf
\supit{e} Boston Micromachines Corporation, Cambridge, MA;
\skiplinehalf
\supit{f} Lawrence Livermore National Laboratory, Livermore, CA
}

\begin{document} 
\maketitle

%%%%%%%%%%%%%%%%%%%%%%%%%%%%%%%%%%%%%%%%%%%%%%%%%%%%%%%%%
\begin{abstract}
Micro-electro-mechanical systems (MEMS) technology can provide for deformable mirrors (DMs) with excellent performance within a favorable economy of scale.  Large MEMS-based astronomical adaptive optics (AO) systems such as the Gemini Planet Imager are coming on-line soon.  As MEMS DM end-users, we discuss our decade of practice with the micromirrors, from inspecting and characterizing devices to evaluating their performance in the lab.  We also show MEMS wavefront correction on-sky with the ``Villages'' AO system on a 1-m telescope, including open-loop control and visible-light imaging.  Our work demonstrates the maturity of MEMS technology for astronomical adaptive optics.
\end{abstract}
%%%%%%%%%%%%%%%%%%%%%%%%%%%%%%%%%%%%%%%%%%%%%%%%%%%%%%%%%%%%%

%%%%%%%%%%%%%%%%%%%%%%%%%%%%%%%%%%%%%%%%%%%%%%%%%%%%%%%%%%%%%
\section{INTRODUCTION}
%%%%%%%%%%%%%%%%%%%%%%%%%%%%%%%%%%%%%%%%%%%%%%%%%%%%%%%%%%%%%

Discoveries in astronomy continue to be driven by developments in technology.  Progress in adaptive optics (AO) is aided by the development of micro-electro-mechanical systems (MEMS) technology for deformable mirrors.  MEMS deformable mirrors (DMs) are fabricated using practices of the semiconductor industry, and therefore present the opportunity for growth through economies of scale.  These micromirrors are being implemented as wavefront-correctors for AO imaging systems from astronomy to vision science.

Because MEMS-technology deformable mirrors are new, extensive laboratory testing was required as part of the development process.  The Center for Adaptive Optics (CfAO) at the University of California--Santa Cruz is heavily involved in MEMS adaptive optics research and development.  We have pursued a ten-year project to develop, test, and implement MEMS AO on sky.  This paper summarizes our efforts and presents our results, including best practices and lessons learned.

%%%%%%%%%%%%%%%%%%%%%%%%%%%%%%%%%%%%%%%%%%%%%%%%%%%%%%%%%%%%%
\subsection{Background}

The CfAO was founded as a National Science Foundation (NSF) Science and Technology Center from 1999--2010, and is now a multi-campus research unit of the University of California (UC).  Its mission is to advance and disseminate the technology of adaptive optics in service to science, health care, industry, and education.  MEMS deformable mirrors play a key role in three of the four CfAO Themes: AO for Extremely Large Telescopes, Extreme Adaptive Optics, and AO for Vision Science.

In order to develop MEMS technology for AO, the CfAO and Boston Micromachines Corporation (BMC) undertook a ten-year project aimed at producing a 4096-element DM for high-order ``extreme'' AO (ExAO) in 2000\cite{olivier2000}.  This project has been fulfilled as specified\cite{norton2009}, and the 4k-MEMS has been installed in an ExAO instrument, the Gemini Planet Imager (GPI).  Here we recount the process of MEMS development from our perspective as end-users in the Laboratory for Adaptive Optics.

%%%%%%%%%%%%%%%%%%%%%%%%%%%%%%%%%%%%%%%%%%%%%%%%%%%%%%%%%%%%%
\subsection{The Laboratory for Adaptive Optics}

The CfAO's Laboratory for Adaptive Optics (LAO) was initiated by the Gordon \& Betty Moore Foundation with \$9M for 6 years, and is currently funded through myriad grants and projects.  The purposes of the LAO are to develop AO techniques for extremely-large ground-based telescopes, to develop and build a planet-finder instrument using ExAO techniques, to test and evaluate new components and technologies as they become available, and to provide a laboratory where students and postdocs are trained in AO hardware and software.  The experimental objectives toward these goals are the testing and validation of new AO concepts (\textit{e.g.}\ ExAO, multi-conjugate/MCAO, and multi-object/MOAO), testing MEMS deformable mirrors, and testing alternative wavefront sensor concepts\cite{cfao2009}.  Because of our sub-nanometer precision instrumentation and our unique position to fulfill the CfAO's MEMS development goals, the LAO has become a leader in MEMS characterization and operation.

\begin{figure}[h]
\begin{center}
	\includegraphics[width=9cm]{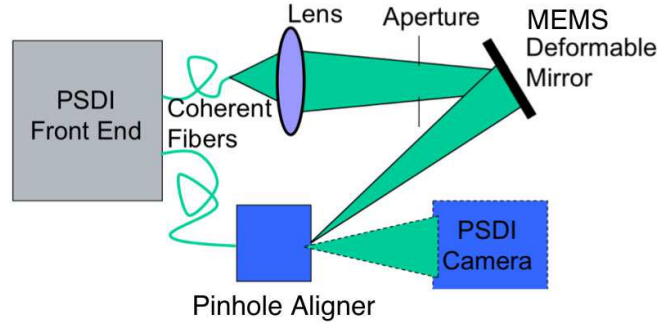}
\end{center}
	\caption{\label{fig:psdi}
	Phase-shifting diffraction interferometer (PSDI) on the ExAO testbed.
	}
\end{figure}

The ExAO testbed at the LAO includes legacy equipment from extreme-ultraviolet lithography work at Lawrence Livermore National Laboratory (LLNL), in the form of an 100-pm absolute accuracy Phase-Shifting Diffraction Interferometer (PSDI).  The PSDI instrument, shown in Fig.~\ref{fig:psdi}, allows us to measure the surface of MEMS devices with sub-nanometer resolution in the phase direction ($z$, along the optical axis).  This is a unique capability and enables us to precisely measure fine positioning of the MEMS surface by its electrostatic actuators.

The goals of the ExAO testbed have been largely met.  These were to demonstrate nanometer metrology of an optical system, to demonstrate $<2$ nm rms wavefront error (WFE), to demonstrate $10^{-7}$ contrast at visible wavelengths and consistency with wavefront measurements\cite{evans2006wavefront}, to install a MEMS deformable mirror, to demonstrate ability to flatten the MEMS to $<1$ nm rms WFE over controllable spatial frequencies\cite{evans2006flat}, to demonstrate contrast of $10^{-6}$ (goal: $10{^-7}$) with a MEMS\cite{evans2006flat}, to characterize the MEMS device and identify performance limitations\cite{evans2005spie, evans2007, morzinski2006, morzinski2009, norton2009, thomas2009}, to demonstrate nanometer-accuracy measurements with a Shack-Hartmann WFS, and testing of other ExAO instrument components\cite{severson2006, evans2009}.  Selected readings for those interested in MEMS performance in the lab are: Evans \textit{et al.}\cite{evans2006flat} describe achieving 0.54 nm rms flatness; Morzinski \textit{et al.}\cite{morzinski2006} describe stability and position repeatability; Norton's work on hysteresis is reported in Morzinski \textit{et al.}\cite{morzinski2008}; and Norton \textit{et al.}\cite{norton2009} characterize the 4k-MEMS for GPI.

%%%%%%%%%%%%%%%%%%%%%%%%%%%%%%%%%%%%%%%%%%%%%%%%%%%%%%%%%%%%%
\subsection{MEMS development}

Economy of scale is a major stimulus for choosing MEMS technology for creation of next-generation deformable mirrors.  Directly imaging planets around other stars requires a high-order correction with an extended spatial cutoff frequency---thus requiring a DM with a high actuator count.  Conventional deformable mirrors have hundreds of degrees of freedom (DoF); MEMS technology allows for thousands of actuators at a fraction of the cost per degree of freedom.  The planet-imaging case is not the only AO regime aided by MEMS DMs, but has been the main driver of the MEMS experiments conducted on the ExAO testbed in the LAO.  MEMS are also ideal for precision open-loop correction required by MOAO; for size- and mass-limited applications such as space-based telescopes; and for cost-conscious portable applications such as human vision testing and diagnosis.

\begin{figure}[h]
\begin{center}
	\includegraphics[width=6cm]{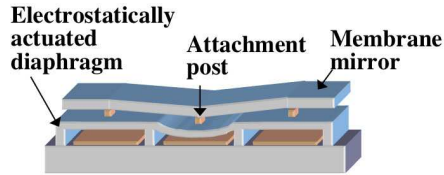}
\end{center}
	\caption{\label{fig:actuator}
	Schematic of 3 MEMS actuators from Boston Micromachines Corp.
	}
\end{figure}

MEMS DMs are fabricated by BMC with surface micromachining of polysilicon thin films\cite{bifano2000}.  Conductive and insulating layers are deposited and selectively etched to build up the actuator structure, shown in Fig.~\ref{fig:actuator}.  The actuators are individually addressable with channels wire-bonded to a ceramic carrier.  A potential is applied to the electrode and its plate is attracted by the electrostatic force; the mirror surface is attached to the plate with posts and deflects according to the square of the voltage.  A thin reflective coating of gold is applied to the facesheet.  The processes and foundries used in MEMS fabrication are shared with the silicon computer industry, allowing for costs to be greatly reduced as compared to DMs made of glass with stiff piezo-electrostrictive actuators.

As with any new technology, the question from an end-user's perspective is, How well do MEMS deformable mirrors work?  This question encompasses stability, environment, operation, modeling, repeatability, dynamic range, spatio-temporal response, and more.  At the CfAO and LAO, we have marked a decade of MEMS research.  Our studies include experimental and theoretical work, both in the laboratory and at the observatory.  Here, we review our MEMS development work from our point of view as MEMS AO users.  We describe installation and characterization of MEMS devices.  We review our laboratory results, and demonstrate MEMS-based AO on-sky at a professional astronomical observatory.  Additionally, we share our customs for care and handling of a MEMS deformable mirror.

%%%%%%%%%%%%%%%%%%%%%%%%%%%%%%%%%%%%%%%%%%%%%%%%%%%%%%%%%%%%%
\section{PROCEEDING TOWARD ON-SKY MEMS ADAPTIVE OPTICS}
%%%%%%%%%%%%%%%%%%%%%%%%%%%%%%%%%%%%%%%%%%%%%%%%%%%%%%%%%%%%%

Over the years, the LAO received approximately ten 1024-actuator MEMS mirrors, and a handful of 144- and 4096-actuator DMs from BMC.  Each mirror was inspected and tested according to a regular procedure.  Our feedback allowed BMC to improve their fabrication processes, and we saw steady improvement in the quality of the devices.  We developed calibration and verification tests for operating the MEMS devices on the ExAO testbed.  One result of this work was, ``Villages,'' a MEMS-based visible-light AO system with open-loop control on a 1-m telescope at Lick Observatory.  In this section we discuss the steps taken to bring a MEMS from initial specification and delivery, to laboratory inspection and testing, to operation in closed and open loop using natural guide stars on a small telescope.

%%%%%%%%%%%%%%%%%%%%%%%%%%%%%%%%%%%%%%%%%%%%%%%%%%%%%%%%%%%%%
\subsection{Specifying the MEMS}

In building an AO system, hardware cannot be acquired without specifying the device requirements.  Development of the specifications for the 4k-MEMS was driven both by the needs of ExAO and of AO for extremely-large telescopes.  The Gemini Planet Imager is the CfAO's and LAO's flagship ExAO instrument\cite{macintosh2006}.  The parameters for GPI were flowed down using a requirement-driven design approach.

\begin{figure}[h]
\begin{center}
	\includegraphics[width=12cm]{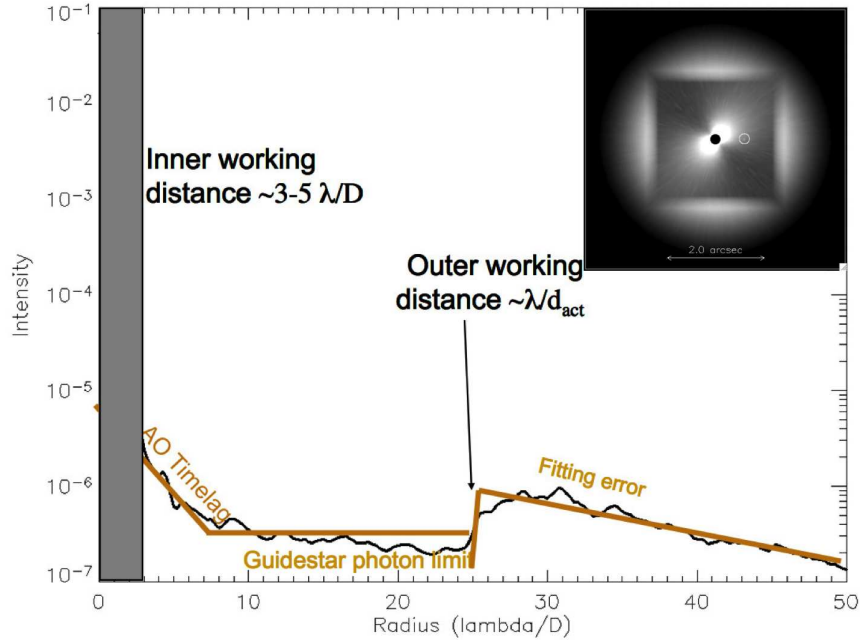}
\end{center}
	\caption{\label{fig:darkhole}
	The ExAO PSF with the dark hole.
	}
\end{figure}

The mission of GPI is to illuminate planetology by conducting a census of hundreds of nearby stars for massive planets on wider orbits.  GPI will determine the frequency and characteristics of planetary systems containing young Jupiter analogs, and will characterize the planets' compositions and structures.  Directly imaging these Jupiter-like planets requires high-order AO to minimize wavefront error and to expand the search space.  The canonical ExAO point-spread function (PSF) is shown in Fig.~\ref{fig:darkhole}.  This high-contrast PSF is characterized by a ``dark hole'' where faint planets can be detected, made possible by removal of diffracted starlight via the coronagraph and AO system.

The depth or contrast of the dark hole dictates the brightness and lower mass limits on the planets that GPI detects.  Assuming aliasing, fitting, and bandwidth errors are well-controlled in the mid spatial frequencies, the floor of the dark hole is set by measurement error in the AO system.  Thus, ExAO requires bright natural guide stars to minimize measurement error.

The spatial extent of the dark hole dictates the widest orbital radii that can be sampled by GPI.  The edge of the dark hole is set by the spatial frequency cutoff of the AO system.  This is the driver for having thousands of degrees of freedom in the deformable mirror, leading to the selection of MEMS technology for ExAO.

The full specifications of the GPI 4k-MEMS include DoF, surface quality, stroke, yield, size, bandwidth, temperature, and more\cite{norton2009}.  The requirements were arrived at through a process including modeling of planetary systems and their detectability; end-to-end simulations of GPI; testing of current BMC MEMS devices; and iterative feedback with BMC.  Boston Micromachines Corp.\ regularly produced 144- and 1024-element DMs; extension to a 4096-element DM was expected to be a straightforward scaling-up of the process.  However, purity and other processing issues became more difficult at this scale.  Therefore it took time, but the final GPI MEMS meets all specifications but yield, the latter being mitigated with optical solutions.  The final science-grade 4k MEMS is now installed in GPI at UC--Santa Cruz for integration and testing.  Norton \textit{et al.}\cite{norton2009} describe the 4k MEMS for GPI.

%%%%%%%%%%%%%%%%%%%%%%%%%%%%%%%%%%%%%%%%%%%%%%%%%%%%%%%%%%%%%
\subsection{Inspecting the MEMS}

\begin{figure}[h]
\begin{center}
	\includegraphics[width=15cm]{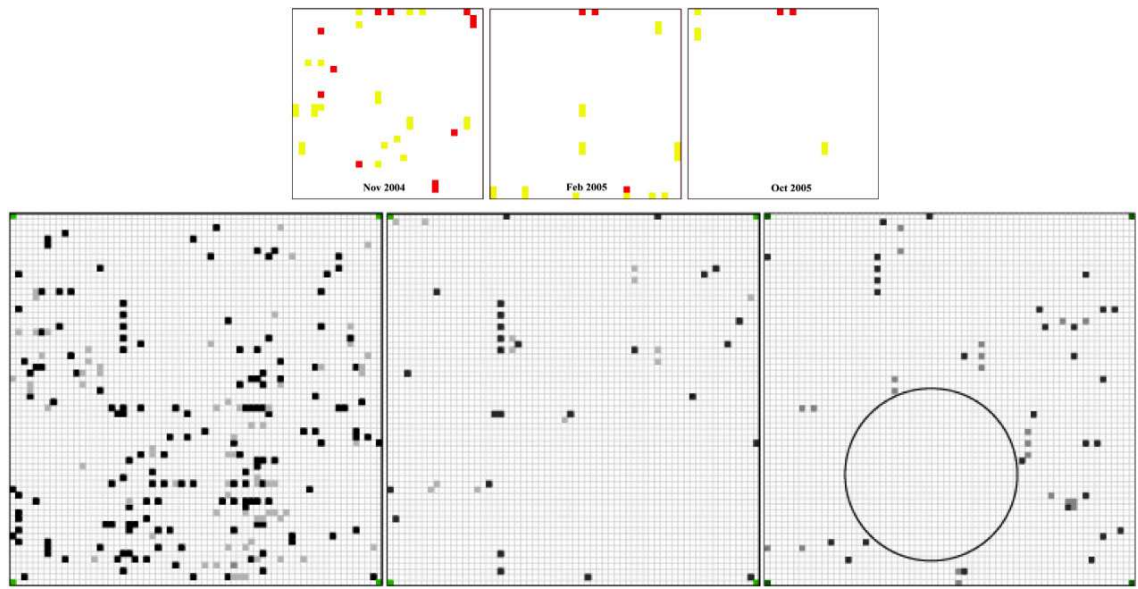}
\end{center}
	\caption{\label{fig:yield}
	MEMS yield improvements over time.
	Top: 1k-MEMS yield improved from 97.9\% to 98.2\% to 99.5\%\cite{evans2006flat} --- white actuators are fully responsive, yellow are semi-responsive, and red are non-responsive.  Bottom: 4k-MEMS yield improved from 94.4\% to 98.8\% -- 99.0\%\cite{norton2009} --- white actuators are fully responsive, gray are semi-responsive, and black are non-responsive.
	}
\end{figure}

Throughout our development project, we regularly received shipment of 1024-actuator MEMS devices from BMC.  Our first step was to perform a visual inspection, by eye and often with a microscope.  The next step was to verify the mapping of the actuators and test their response to voltage.  We conduct a ``rows-and-columns'' test to do this, applying 160 V to each row and then each column sequentially.  We measure the surface with the PSDI on the ExAO testbed, looking for mis-mapped or irregular actuators.  Subtracting the 0-volt image from the 160-volt image shows dead actuators.  Actuators can also be over- or under-responsive, coupled to a neighbor, or responsive only to the level of the bias.  The yield of fully-responsive actuators improved over the years to 99.5\% for a 1k-MEMS and 99.0\% for a 4k-MEMS (Fig.~\ref{fig:yield}).

Irregular actuators are characterized in detail by calibrating the deflection as a function of voltage for each actuator.  Because the influence function falls to zero at a spacing of a couple actuators, each actuator can be measured individually by applying a voltage to every fourth actuator across a row and down a column.  This allows the test to be done for all 1024 or 4096 actuators in four sets.  We step from the minimum to maximum voltage (\textit{e.g.}\ 0 to 200 V) in 20-V increments, with the other actuators set to a bias of the mid-displacement voltage for a single actuator (\textit{e.g.}\ 140 V).

%%%%%%%%%%%%%%%%%%%%%%%%%%%%%%%%%%%%%%%%%%%%%%%%%%%%%%%%%%%%%
\subsection{Calibrating the MEMS}

\begin{figure}[h]
\begin{center}
	\includegraphics[width=9cm]{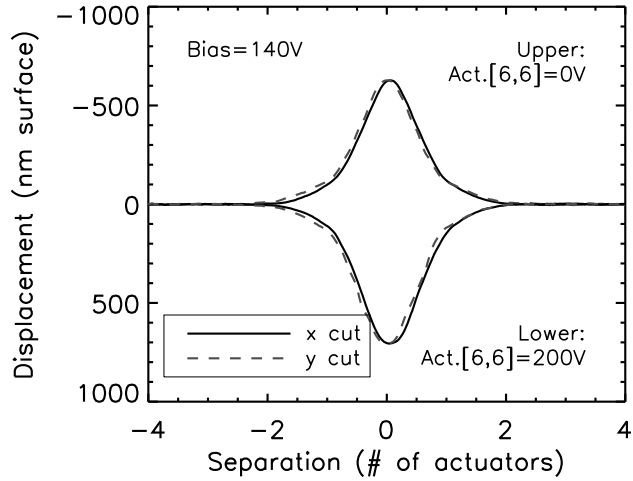}
\end{center}
	\caption{\label{fig:inffxn}
	Influence function, Villages 144-MEMS.
	}
\end{figure}

To shape the wavefront, the MEMS must be calibrated for its voltage response.  The influence function is the displacement of the entire MEMS surface when a voltage is applied to a single actuator.  Because the MEMS has a stiff continuous facesheet, the neighboring actuators are displaced somewhat as well.  Figure~\ref{fig:inffxn} shows the influence function of the 144-MEMS used in Villages.  The single actuator influence function for a 1k-MEMS falls to 26\% at one actuator distance and 4\% at two actuators distance, and is uniform with varying bias voltage and applied voltage.

\begin{figure}[h]
\begin{center}
	\includegraphics[width=9cm]{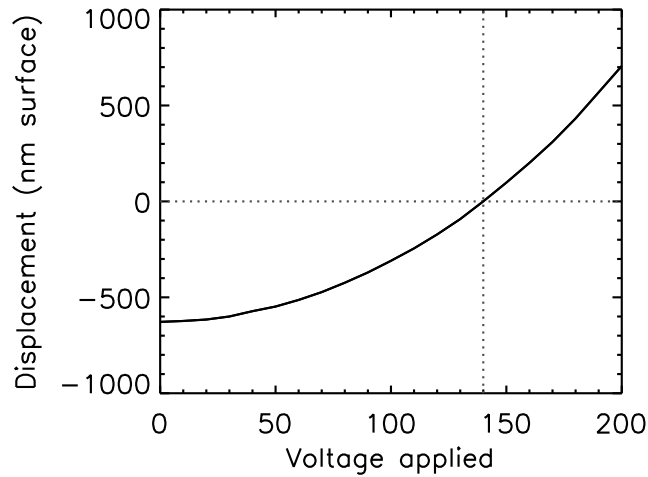}
\end{center}
	\caption{\label{fig:voltdisp}
	Displacement vs.\ voltage, single actuator, Villages MEMS.
	}
\end{figure}

The displacement as a function of voltage is needed for applying shapes.  The voltage that gives the displacement in the middle of the range is used as the bias.  Figure~\ref{fig:voltdisp} shows the single-actuator displacement curve for the 144-actuator Villages MEMS.  Stroke has improved with advanced designs, after trading with facesheet thickness, surface curvature, and actuator spacing.  The GPI 4k-MEMS has 1 $\mu$m of unpowered curvature, which will be removed by the woofer, in a trade for its excellent surface quality of 5.74 nm surface rms\cite{norton2009}.

\begin{figure}[h]
\begin{center}
	\includegraphics[width=8cm,angle=90]{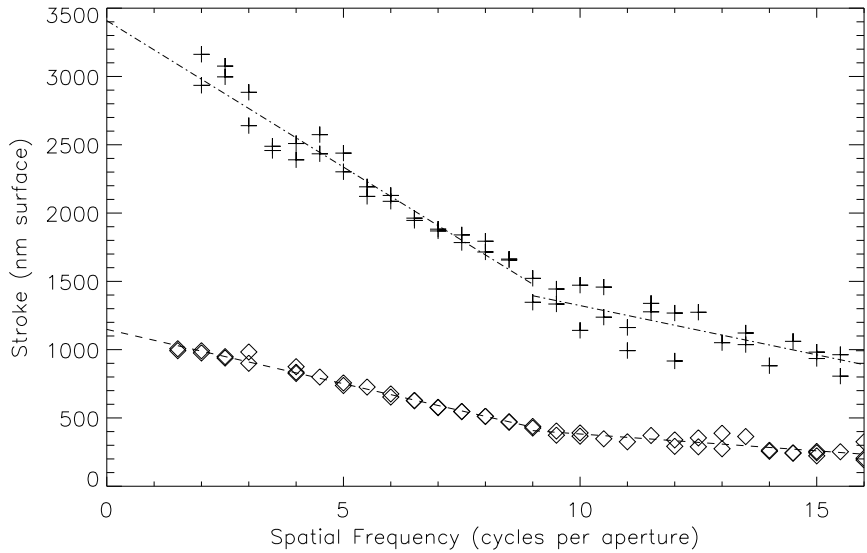}
\end{center}
	\caption{\label{fig:transferfxn}
	Transfer function: Stroke vs.\ spatial frequency\cite{morzinski2009}.
	Diamonds: 340-$\mu$m pitch device, operated from 0--160 V.
	Plusses: 400-$\mu$m pitch device, operated from 0--200 V.
	}
\end{figure}

The stroke should be measured to ensure it is adequate for the performance requirements.  The voltage-displacement curve gives the maximum stroke achieved for a single actuator.  However, because of the broad influence function, stroke varies with shape.  Applying 1-d sinusoids at the controllable frequencies the MEMS (from 1 to 16 waves across the device for the 32x32 MEMS) allows the stroke to be measured as a function of frequency.  Figure~\ref{fig:transferfxn} shows the result, which is the transfer function of the MEMS and gives the power that is corrected at each spatial frequency\cite{morzinski2009}.  The trade for stroke with actuator spacing and voltage applied can also be seen.

If the MEMS is to be operated in open loop control, the phase-to-volts model must be calibrated.  This process is used to populate a look-up table in which the correct voltage to apply to each actuator is found by searching for the desired displacement.  Details are in Morzinski \textit{et al.}\cite{morzinski2007}, and the process entails measuring the permutations of all voltages applied to a central actuator and to the neighbors, in order to separate the spring force from the electrostatic force.  Successful MEMS open loop models have also been produced by Stewart \textit{et al.}\cite{stewart2007} and Blain \textit{et al.}\cite{blain2000}.

The calibrations described above require an interferometer, but a commercial instrument such as a Zygo can be used; the PSDI is not necessary for these tests as sub-nanometer precision is not required.

%%%%%%%%%%%%%%%%%%%%%%%%%%%%%%%%%%%%%%%%%%%%%%%%%%%%%%%%%%%%%
\subsection{Testing the MEMS}

\begin{figure}[h]
\begin{center}
	\includegraphics[width=15cm]{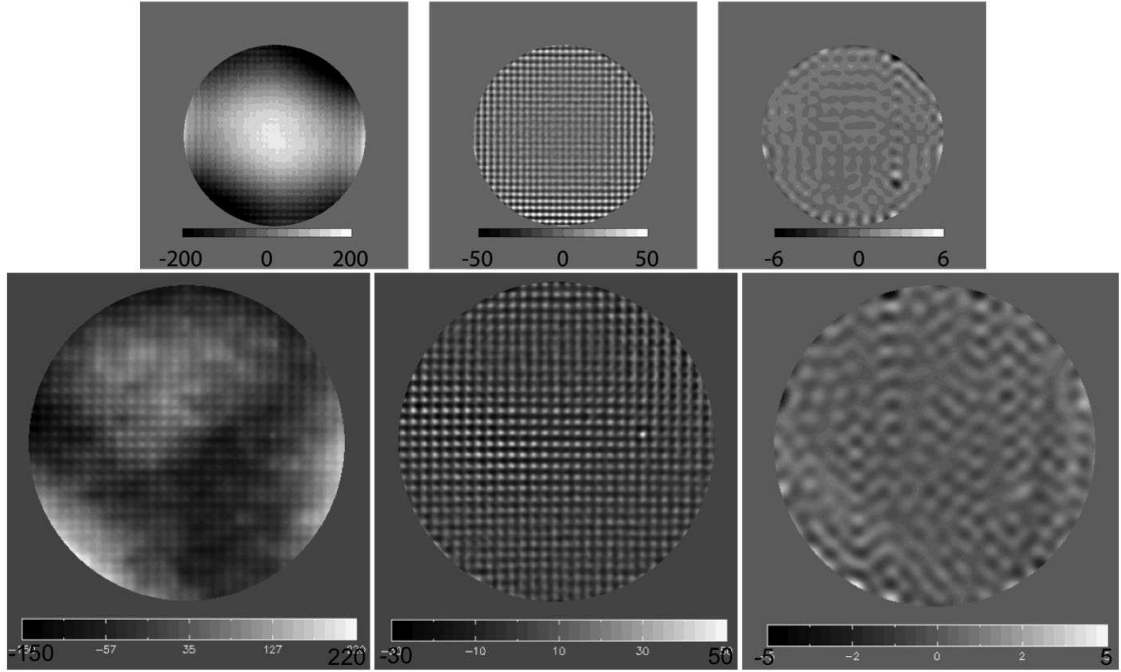}
\end{center}
	\caption{\label{fig:flatmems}
	Best MEMS flattening.  Top: 1k-MEMS\cite{evans2006flat}.  Bottom: 4k-MEMS\cite{norton2009}.
	Left: Unpowered shape (148.1 and 47.6 nm phase rms, respectively).  Center: flat shape, all spatial frequencies (12.8 and 10.3 nm phase rms, respectively).  Right: flat shape, controllable spatial frequencies (0.54 and 0.51 nm phase rms in-band, respectively).
	}
\end{figure}

Once the MEMS is inspected and calibrated, it is ready to be flattened in the lab as a test of wavefront control.  We flattened a 1k-MEMS to 0.54 nm phase rms in-band and the 4k-MEMS to 0.51 nm phase rms in-band.  Figure~\ref{fig:flatmems} shows the results of MEMS flattening with PSDI on the ExAO testbed.  Details are in Evans \textit{et al.}\cite{evans2006flat} and Norton \textit{et al.}\cite{norton2009}.

Other tests we conducted include stability\cite{morzinski2006}, position repeatability\cite{morzinski2006}, and hysteresis\cite{morzinski2008}, all of which are satisfactory to less than a nanometer.  Because the MEMS is stable and repeatable, we find it useful to save the voltage commands that flatten the device; for example, the flattening voltages are applied to the Villages MEMS when aligning the system and image sharpening to as to ensure a flat DM.

The PSDI is required for sub-nanometer measurements, as when flattening to $<1$ nm and characterizing the stability and go-to capability.

%%%%%%%%%%%%%%%%%%%%%%%%%%%%%%%%%%%%%%%%%%%%%%%%%%%%%%%%%%%%%
\subsection{Operating the MEMS}

\begin{figure}[h]
\begin{center}
	\includegraphics[width=9cm]{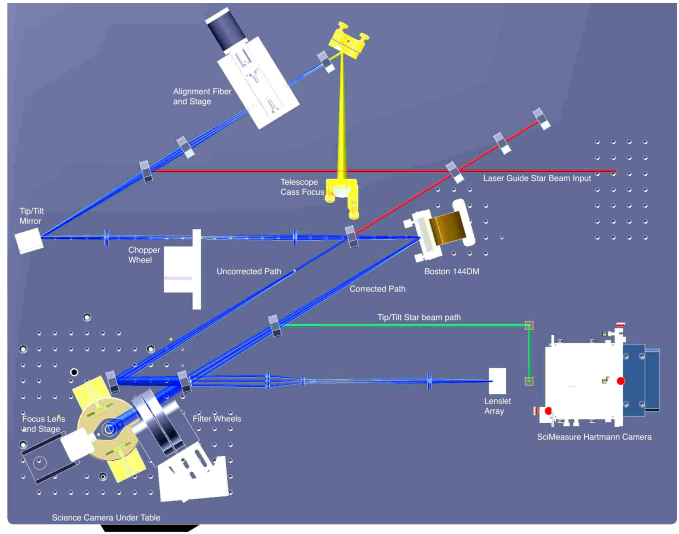}
\end{center}
	\caption{\label{fig:villageslayout}
	Layout of the Villages AO system.
	}
\end{figure}

\begin{figure}[h]
\begin{center}
	\includegraphics[width=5cm]{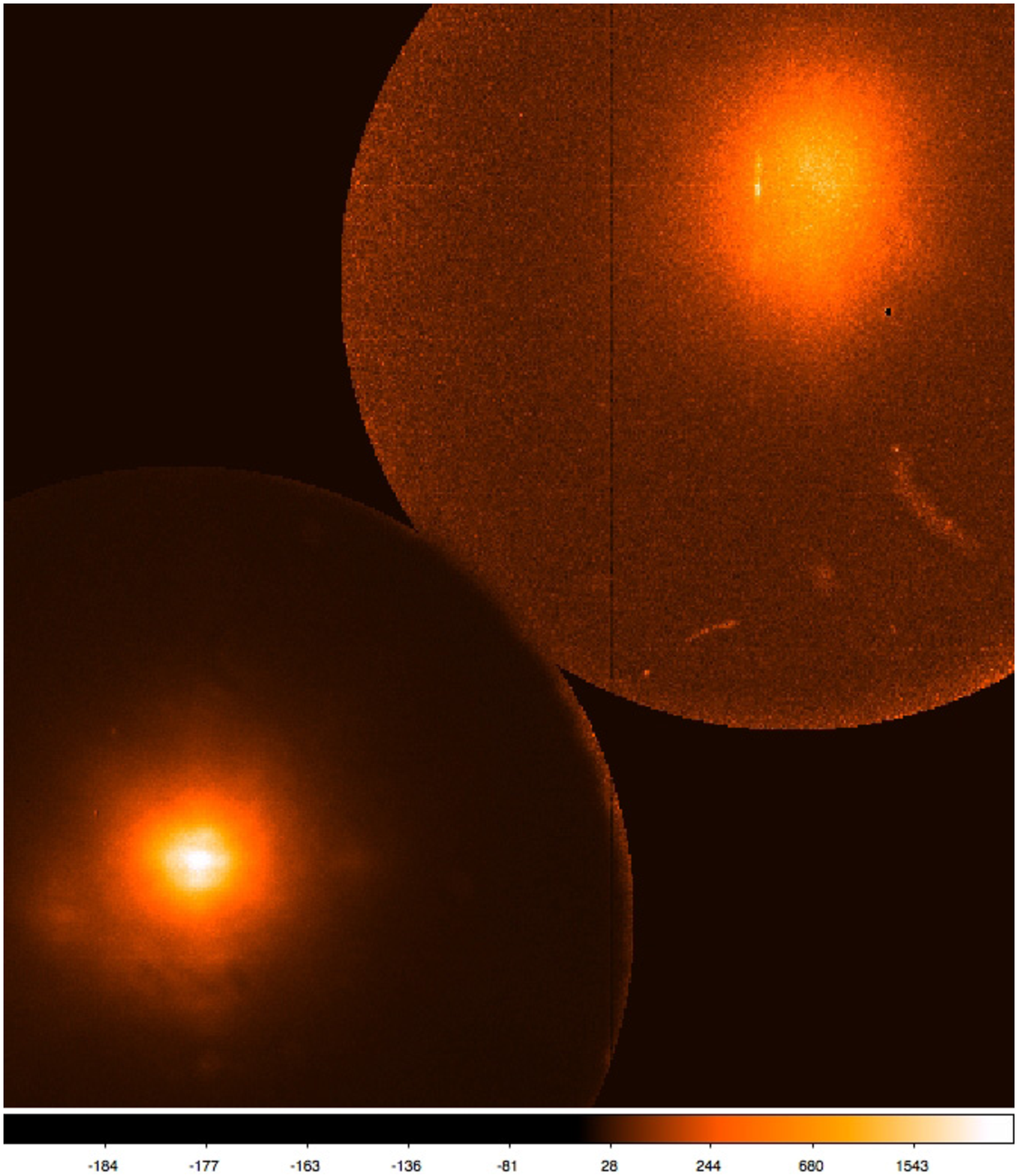}
\end{center}
	\caption{\label{fig:villagespsf}
	Corrected (lower left) and uncorrected (upper right) on-sky natural guide star image on the Villages science camera, log scale.
	}
\end{figure}

The final step is operating the MEMS in an on-sky AO system.  The Villages AO system can be thought of as an on-sky testbed for MEMS AO, open-loop control, and visible-light AO.  It has been located on the 1-m ``Nickel'' telescope at Lick Observatory since Fall 2007.  Its layout is shown in Fig.~\ref{fig:villageslayout}; Villages has two optical paths for a side-by-side comparison of corrected and uncorrected PSFs on the science camera (Fig.~\ref{fig:villagespsf}), and closed and open loop Shack-Hartmann spots on the wavefront sensor camera.

\begin{figure}[h]
\begin{center}
	\includegraphics[width=11cm]{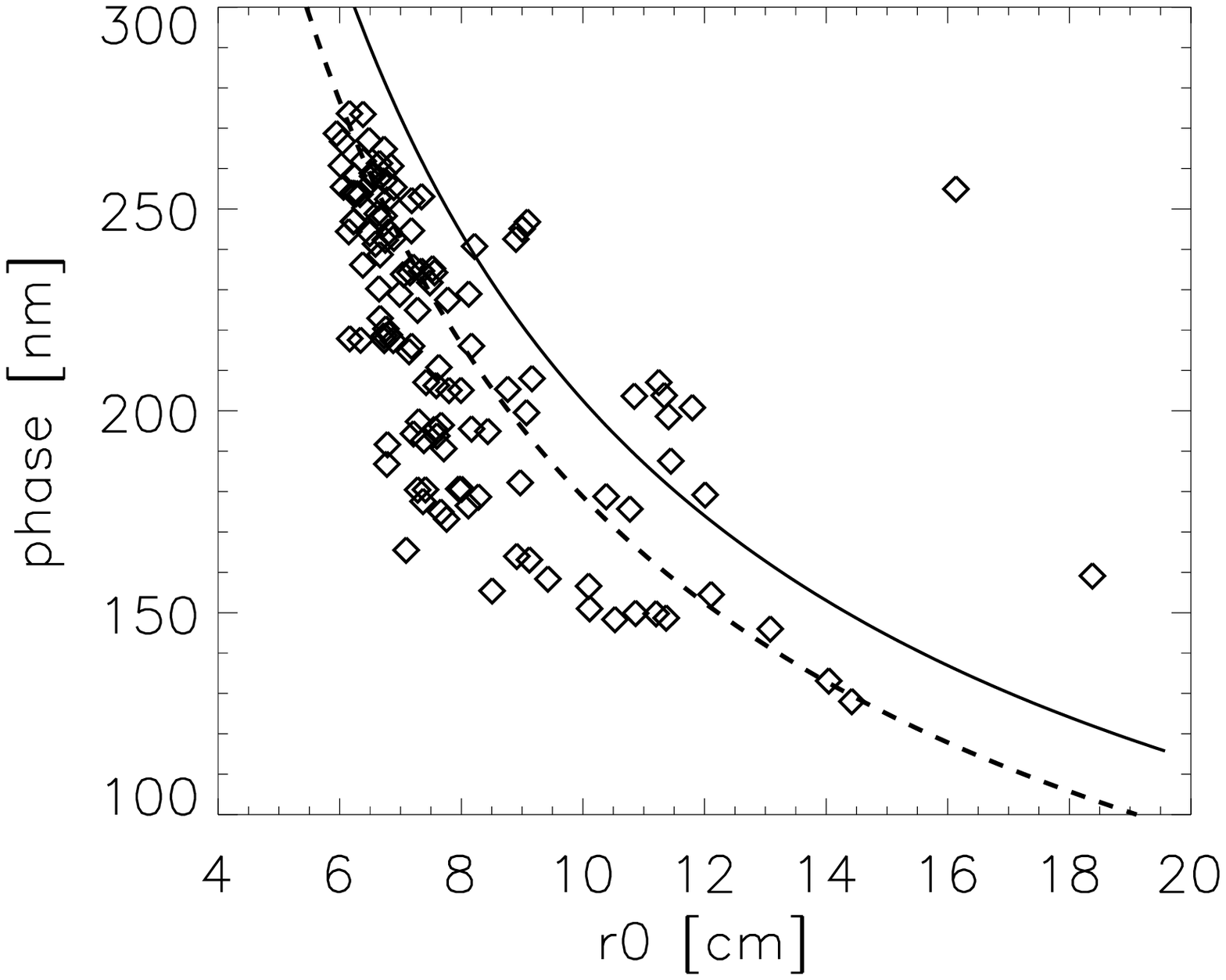}
\end{center}
	\caption{\label{fig:villagesresults}
	Villages results, showing residual WFE as a function of input wavefront for all images from the Sep.\ 2010 run.
	$r_0$ is measured from the FWHM of the uncorrected images.  The residual phase is determined from the Marechal approximation.
	Vibration tests and other experiments are included, which accounts for the bad data; the lower locus of points are for typical AO conditions.
	The solid line is the analytic form for Kolmogorov turbulence with piston and tip-tilt removed; data points above this line are likely affected by vibration or other issues.
	The dashed line is a best fit to the good data using a two-parameter fit representing an AO system which removes 10\% plus a constant 40 nm WFE.
	}
\end{figure}

Villages' unique optical layout allows for side-by-side comparison of AO control and effects.  Figure~\ref{fig:villagesresults} summarizes the AO control achieved on all closed-loop images for the run in Sep.\ 2010.  The $x$-axis plots $r_0$ measured from the full-width at half-max (FWHM) of the uncorrected images, using the formula FWHM = $\lambda/r_0$.  The $y$-axis plots the residual WFE after closed-loop control of the corrected images, using the extended Marechal approximation $S = \exp{(-\sigma^2)}$.  All data points are plotted, even those taken during gain adjustments and during experiments with motors in the dome to search for a vibration source.  Therefore, we take the lower locus of points as results for typical AO conditions.  Then we see that for typical seeing conditions of $r_0 \approx 10$cm, the residual WFE was 150 nm phase rms in closed-loop.

After about 4 years and 1000 hours of operation, the 144-actuator Villages MEMS is shown to be robust.  The flattening voltages continue to give a sharp internal PSF, and no actuators have broken.  The device has a window to protect against humidity-induced oxidation, and it is powered down when not in use.  The MEMS is not a dominant source in the error budget, as reported in Morzinski \textit{et al.}\cite{morzinski2010}.

%%%%%%%%%%%%%%%%%%%%%%%%%%%%%%%%%%%%%%%%%%%%%%%%%%%%%%%%%%%%%
\section{BEST PRACTICES}
%%%%%%%%%%%%%%%%%%%%%%%%%%%%%%%%%%%%%%%%%%%%%%%%%%%%%%%%%%%%%

In this section we share our tips and lessons-learned for MEMS adaptive optics.

%%%%%%%%%%%%%%%%%%%%%%%%%%%%%%%%%%%%%%%%%%%%%%%%%%%%%%%%%%%%%
\subsection{Care and handling of the MEMS}

There are two common failure mechanisms for MEMS actuators: snap down and humidity damage.  In a snap-down failure the actuator has had too much voltage applied and the electrostatic force overcomes the restoring spring force and the actuator gets stuck in the highest-volt position.  It is possible but not recommended that snapped down actuators can be freed by poking them with a probe.  Snap-down is prevented with hardware and software safety stops to the maximum voltage that can be applied.

\begin{figure}[h]
\begin{center}
	\includegraphics[width=6cm]{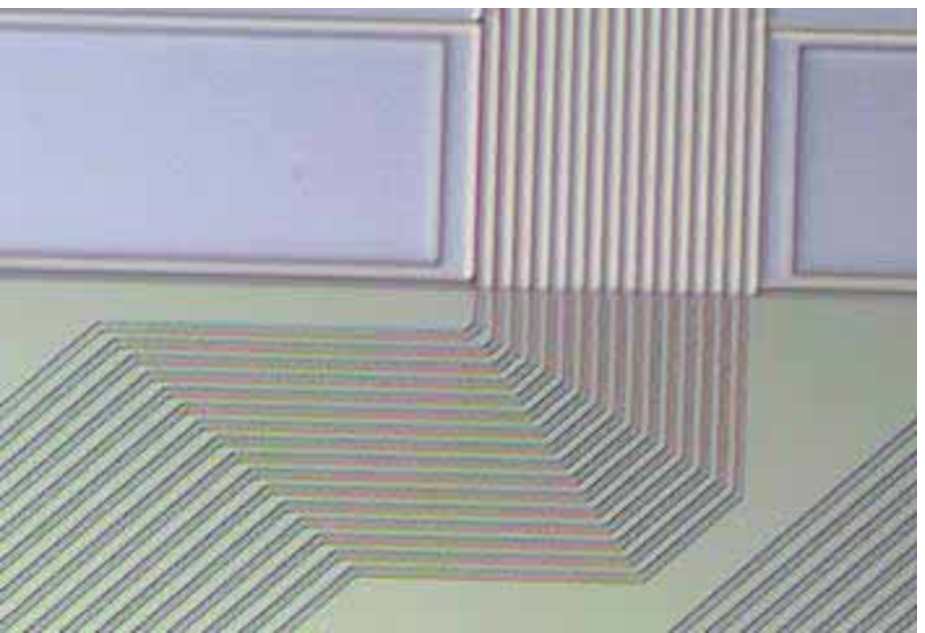}
	\includegraphics[width=6cm]{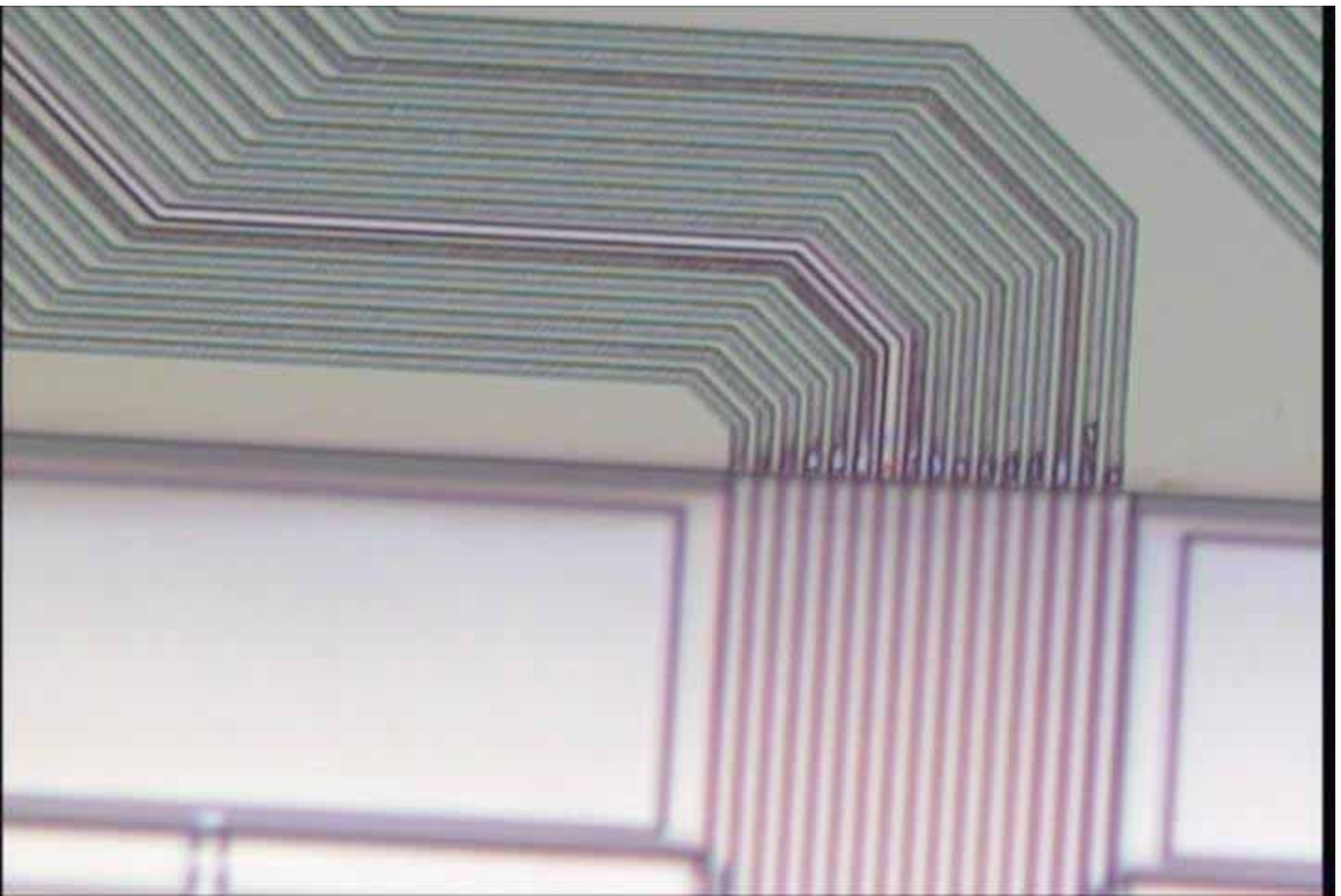}
\end{center}
	\caption{\label{fig:corrosion}
	Humidity corrosion, 2005, photos from BMC.  Left: undamaged.  Right: wire damage.
	}
\end{figure}

Voltage in the presence of humidity can induce dissociation of water, causing anodic oxidation of poly-silicon\cite{shea2000, plass2003, shea2004, hon2007}.  Water dissociates into $OH^-$ and $H^+$; the negatively-charged $OH$ ions are attracted to the positive anode, where they react with the poly-silicon to form non-conducting $SiOH$ and $SiO_2$, also known as glass.  Over time with a combination of high voltage and high humidity, the connections for the actuators oxidize and the actuator gets stuck in the zero-volts position.  It is possible that if there is oxidation on the device it would be visible under a microscope (see Fig.~\ref{fig:corrosion} showing early corrosion during shipping).

We operate our devices at a ``negative'' voltage, such that the surface is kept at a positive voltage (anode) and the wire bond traces are at a negative voltage (cathode).  This gives greater protection to the thin wires from corrosion and prevents short circuiting individual actuators, because any oxidation at the anode is negligible relative to the large size of the common electrode.  Furthermore, any two unpassivated silicon surfaces separated by a distance allowing electric fields above a certain level can attract $OH^-$ ions and oxidize silicon on the positive side, if water is present in the air.  This would also apply to two adjacent traces separated by 2 $\mu$m with widely disparate voltages --- the more positive wire will oxidize if the relative humidity is high enough.

Our laboratory is temperature and humidity controlled to prevent humidity damage.  If the MEMS does not have a window, it should only be operated at voltage when the humidity is $<50$\%.  For on-sky devices, BMC can fit an hermetically-sealed window to keep out moisture.  For greater protection against water leach, GPI will cycle dry nitrogen through the window.

We only had one actuator lose functioning on our watch, and it was on a windowless device.  This was actuator [15,15] on device W10\#X, and it is directly adjacent to a low-response actuator, such that its voltage was always higher than its neighbor, so it is possible it is cumulative oxidation.  It occurred about two years after the device was received, and until the actuator died this was our best device and our workhorse.  The device has no window, but we were quite careful in using it only when relative humidity was $<40$\% (to be safe).  Our subsequent best device, W107\#X, has a window and was used extensively for $>3$ years and remains in good condition.  Thus out of ten 1k-MEMS and a handful of 144-MEMS and 4k-MEMS, only 1 actuator in $\sim$10$^4$ died on our watch.

All other actuators that seemed to die on us were determined, instead, to be problems at the socket or the electronics.  Problems with the electronics were determined by rotating the boards and checking which actuator is dead, and usually remedied by new boards.  Problems at the socket were determined by inspecting the connectors for bent pins with a magnifying glass, and testing the bent pins with an ammeter.  We determined that repeatedly transferring devices from the PSDI to the Zygo interferometer was causing bent pins at the ZIF socket, so we recommend minimizing transfer, as the ZIF is not strictly ``zero-insertion force.''  Additionally, static must be carefully avoided when handling the MEMS.

We had a device early on in testing on which we often left the same 4 actuators (used for alignment) strongly poked in the maximum voltage position with the rest of the device at a lower bias, on occasion for hours to days at a time.  Over time these actuators did not return completely to the zero-volts position, but bumped slightly when unpowered.  This cause still remains unknown, but the actuators remain functional.  We modified our habits to unpower the device when not in use, and this has not been a problem with subsequent devices.

%%%%%%%%%%%%%%%%%%%%%%%%%%%%%%%%%%%%%%%%%%%%%%%%%%%%%%%%%%%%%
\subsection{Tips and tricks}

\begin{figure}[h]
\begin{center}
	\includegraphics[width=5cm]{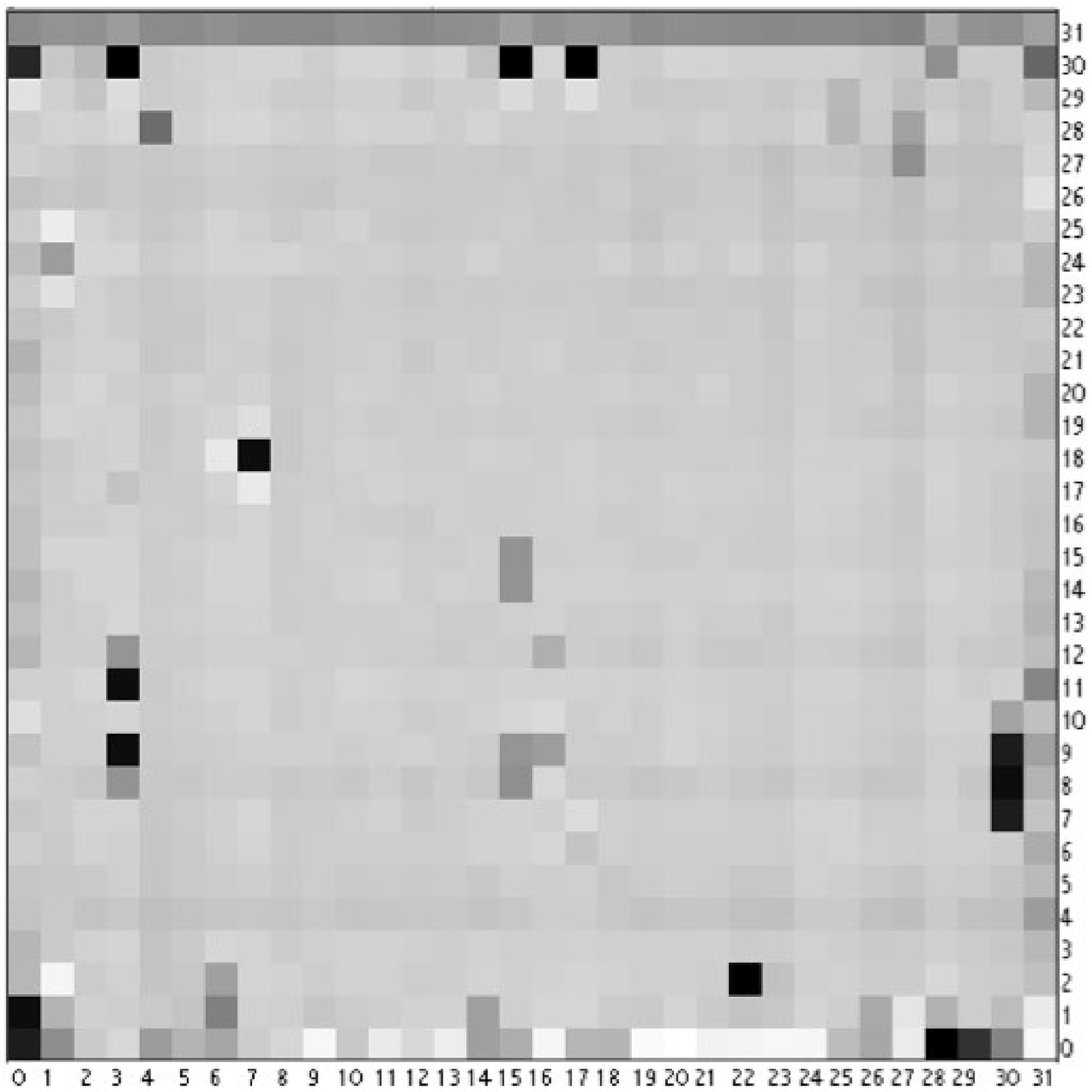}
	\includegraphics[width=9cm]{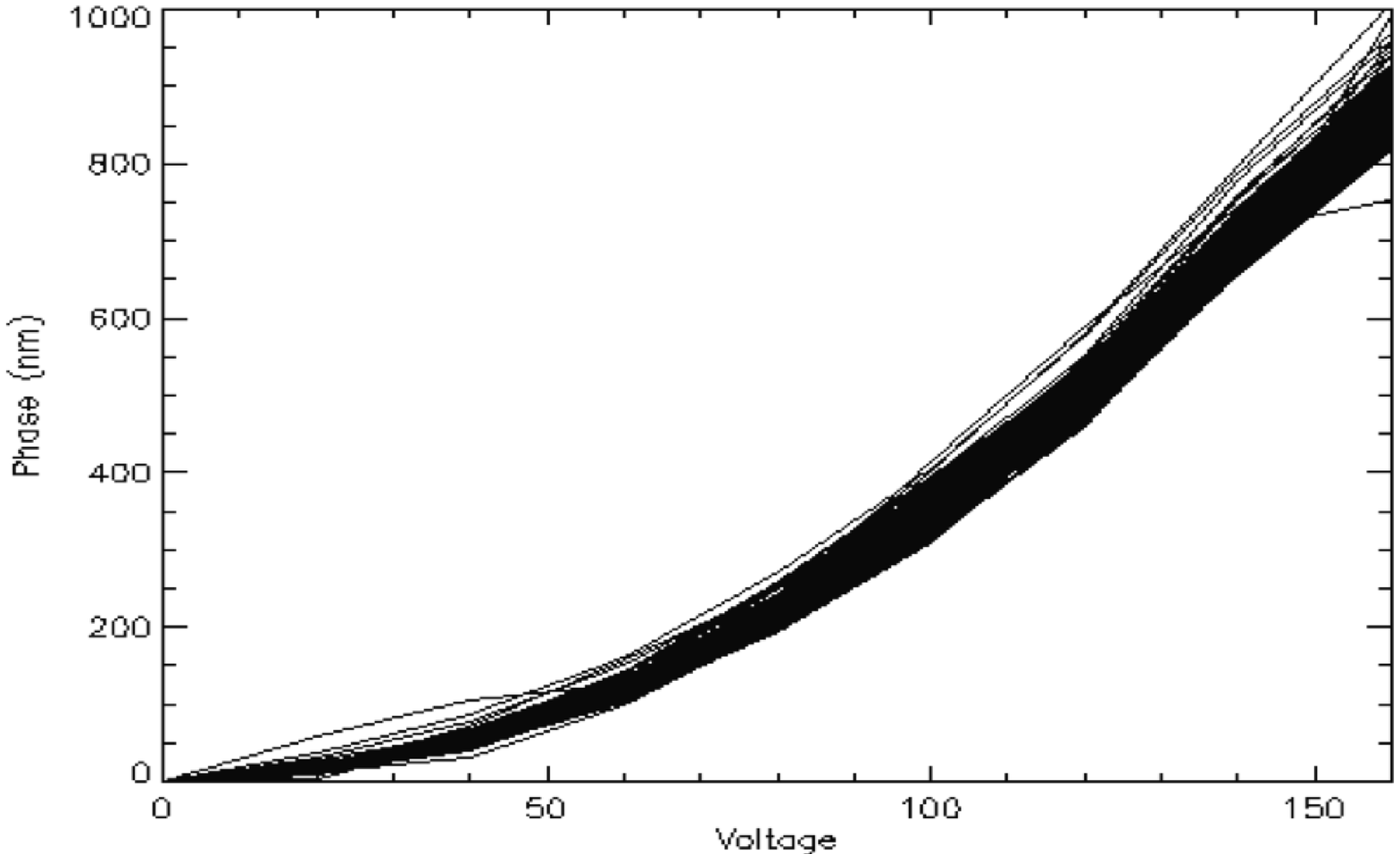}
\end{center}
	\caption{\label{fig:layramaxvolt}
	Left: Displacement at maximum voltage, 1k-MEMS.
	White: over-responsive actuator.  Dark gray: under-responsive actuator.  Black: no-response actuator.
	Right: Individual voltage-displacement curves for the functional actuators.
	}
\end{figure}

When calibrating the voltage-displacement curve (recall Fig.~\ref{fig:voltdisp}), we usually test only four actuators, fit a quadratic, and take the average curve for the entire device.  However, if one takes the time to calibrate all actuators individually, there is an effect.  Figure~\ref{fig:layramaxvolt} shows the displacements of the actuators on a 1k-MEMS at maximum voltage.  Setting aside the dead, weak, and coupled actuators, the fully-functional actuators show a 2--3\% variation in their stroke.  This effect would account for 20--30 nm rms wavefront error in an open-loop control model using an average rather than individual voltage-displacement curve for each actuator.

\begin{figure}[h]
\begin{center}
	\includegraphics[width=9cm]{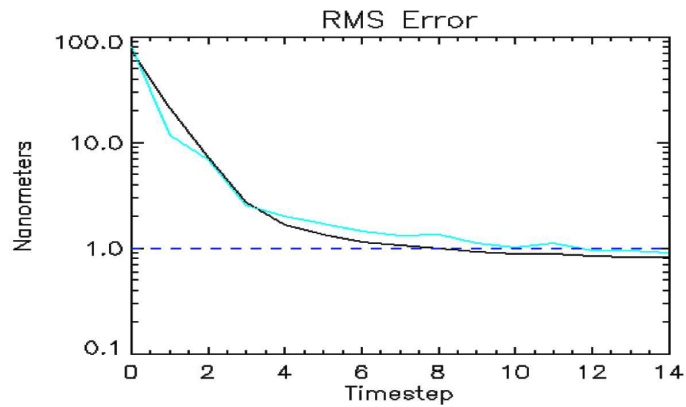}
\end{center}
	\caption{\label{fig:layraclosedloop}
	Closed loop performance: flatness in nm phase rms vs.\ closed-loop iteration.
	Teal: average voltage-displacement curve.  Black: individual voltage-displacement curve.
	Dashed line: goal: $<1$ nm phase rms in-band.
	Using the exact fit for each actuator improves closed loop speed and stability.
	}
\end{figure}

Furthermore, closed-loop performance can be tweaked with an individual voltage-displacement curve for each actuator.  Figure~\ref{fig:layraclosedloop} shows the flattening performance using the average and the individual curves.  The loop converged in only 9 steps as opposed to 13 steps for the case with individual actuator curves, and was a more smooth and stable process.  Thus depending on one's system requirements, it may be worthwhile to consider individually calibrating each actuator.

\begin{figure}[h]
\begin{center}
	\includegraphics[width=12cm]{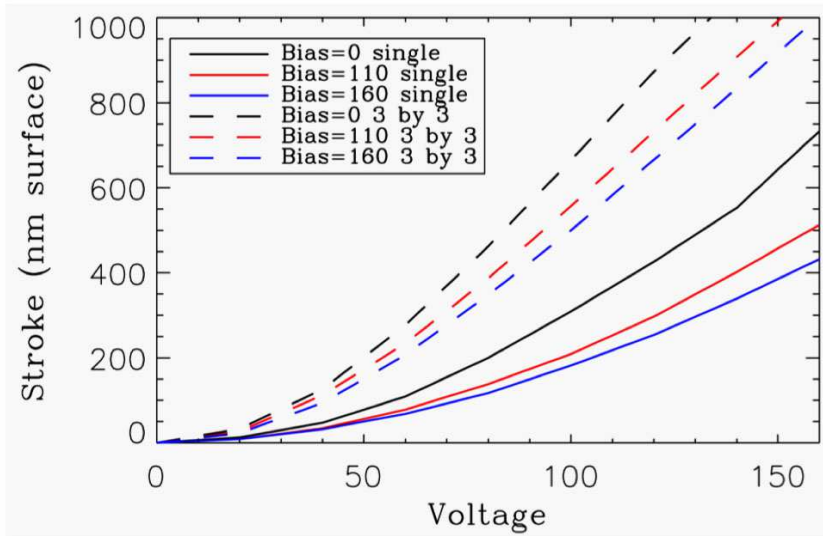}
\end{center}
	\caption{\label{fig:juliastroke}
	Stroke as a function of voltage for a single actuator vs.\ a 3x3 region, and for a bias of 0, 110, and 160 V.
	}
\end{figure}

When specifying the MEMS, it is important to understand the stroke requirements in detail for high-order systems such as ExAO.  Stroke was improved over the years by trading with facesheet thickness, overall curvature and surface quality, and actuator spacing.  We recall from Fig.~\ref{fig:transferfxn} that stroke is a decreasing function of spatial frequency due to the broad influence function caused by the stiff facesheet.  This can also be seen in Fig.~\ref{fig:juliastroke} in which a single-actuator shows less stroke than a 3x3 region of nine actuators.  Interestingly, Fig.~\ref{fig:juliastroke} also shows that stroke is a function of bias, because the electrostatic force is stronger than the restoring spring force.

In order to ensure usage of the device's full stroke range, one must select the proper bias to be in the mid-displacement for the expected operating conditions.  For example, a series of high-spatial-frequency shapes would perhaps be better suited to the single-actuator-stroke bias, whereas shapes with lower spatial frequencies need to be set to a lower bias.  Table~\ref{tab:biasvolts} shows the bias that gives the mid-displacement for a single actuator curve and a 3x3 region, for three different maximum voltages.
\begin{table}[h]
	\centering
	\begin{tabular}{ccc}
		\hline
		 & Mid-displacement & Mid-displacement \\
		Maximum & voltage & voltage \\
		voltage & (single actuator) & (3x3 region) \\
		\hline
		\hline
		160   &   110   &   90   \\
		200   &   140   &   120   \\
		225   &   160   &   140   \\
		\hline
	\end{tabular}
	\caption{
		Suggested bias voltages.
		} \label{tab:biasvolts}
\end{table}

%%%%%%%%%%%%%%%%%%%%%%%%%%%%%%%%%%%%%%%%%%%%%%%%%%%%%%%%%%%%%
\section{CONCLUSIONS}
%%%%%%%%%%%%%%%%%%%%%%%%%%%%%%%%%%%%%%%%%%%%%%%%%%%%%%%%%%%%%

In this paper we have summarized the CfAO development project to produce a 4k-MEMS for the ExAO Gemini Planet Imager and for AO for extremely large telescopes, working with Boston Micromachines Corp.  We have a decade of experience with MEMS DMs in the Laboratory for Adaptive Optics, and almost half a decade at Lick Observatory.  The steps that we took to get a MEMS on-sky were specification, inspection, calibration, testing, and operation.

Laboratory testing demonstrated that MEMS DMs meet requirements for GPI.  On-sky tests confirm reliability in the telescope environment.  The on-sky Villages AO system achieves 150 nm rms WFE under typical seeing on a system in which the subaperture size is approximately equal to the Fried parameter ($d/r_0$$\approx$1).  The 144-MEMS device in Villages has been operated for a total of approximately 1000 hours and remains robust and repeatable, with a stable flat shape and no noticeable change in any of the actuators.

Only one actuator died on us in the lab, out of $10^4$ actuators total on our $\sim$15 MEMS devices.  All other newly-appearing bad actuators were not bad actuators on the MEMS but were bad channels or connectors in the boards or socket.  No actuators went bad at the telescope.  We recommend minimizing device swapping to save the ZIF connectors; to use hardware and software safety stops on the maximum voltage to prevent snap-down; to minimize the humidity or protect the device with a window; and to set the voltages to zero or turn off the power to the MEMS when not in use.

In order to use the full stroke range, it is important to select a bias voltage that is appropriate for the shapes the MEMS will be forming, in order to account for the broad influence function.  Also because of the stiff facesheet, a couple rows of actuators outside the illuminated pupil should be slaved to the neighbors inside the pupil.  Fully-functioning actuators vary by about 2--3\% in displacement response to voltage.  Therefore, closed and open loop performance can be improved with use of individual actuator displacement curves rather than a device average.

We have not seen any problems at the telescope with the Villages MEMS.  Astronomical AO with a MEMS DM is giving 150 nm rms WFE at Lick Observatory with a 144-actuator device on a 1-m telescope.  The MEMS is not a significant term in the Villages error budget.  The 4k-MEMS for GPI meets specification and is currently installed in GPI at UC--Santa Cruz for integration testing.  MEMS technology is robust and reliable, and ready for use in next-generation astronomical AO systems.

%%%%%%%%%%%%%%%%%%%%%%%%%%%%%%%%%%%%%%%%%%%%%%%%%%%%%%%%%%%%%
\section*{ACKNOWLEDGMENTS}
%%%%%%%%%%%%%%%%%%%%%%%%%%%%%%%%%%%%%%%%%%%%%%%%%%%%%%%%%%%%%
This work was performed in part under contract with the Jet Propulsion Laboratory and is funded by NASA through the Sagan Fellowship Program under Prime Contract No.\ NAS7-03001.  JPL is managed for the National Aeronautics Space Administration (NASA) by the California Institute of Technology.
Furthermore, data presented herein were obtained during a Michelson Fellowship supported by the National Aeronautics and Space Administration and administered by the Michelson Science Center.
Any opinions, findings, and conclusions or recommendations expressed in this publication are those of the authors and do not necessarily reflect the views of the National Aeronautics Space Administration (NASA) or of The California Institute of Technology.
Support for this work was also provided by a grant from the Gordon and Betty Moore Foundation to the Regents of the University of California, Santa Cruz, on behalf of the UCO/Lick Laboratory for Adaptive Optics.  The content of the information does not necessarily reflect the position or the policy of the Gordon and Betty Moore Foundation, and no official endorsement should be inferred.
Additionally, support for this work was provided by a minigrant from the Institute of Geophysics and Planetary Physics through Lawrence Livermore National Laboratory.
This research was supported in part by the University of California and National Science Foundation Science and Technology Center for Adaptive Optics, managed by the UC Santa Cruz under cooperative agreement No.~AST-9876783.
Portions of this work were performed under the auspices of the U.~S.~Department of Energy by the University of California, Lawrence Livermore National Laboratory under Contract W-7405-ENG-48.
The Villages experiment was supported through a Small Grant for Exploratory Research from the Astronomy Division of the National Science Foundation, award number 0649261.
Any opinions, findings, and conclusions or recommendations expressed in this publication are those of the authors and do not necessarily reflect the views of the National Science Foundation, the University of California, Boston Micromachines Corporation, Boston University, Lawrence Livermore National Laboratory, or any other entity.

%%%%%%%%%%%%%%%%%%%%%%%%%%%%%%%%%%%%%%%%%%%%%%%%%%%%%%%%%%%%%%%%%%%%%%%%
%%%%%%%%%%%%%%%%  REFERENCES  %%%%%%%%%%%%%%%%%%%%%%%%%%%%%%%%%%%%%%%%%%

\bibliography{spie825303.astroph.bbl}

\begin{thebibliography}{10}

\bibitem{olivier2000}
{Olivier}, S.~S., {Bierden}, P.~A., {Bifano}, T.~G., {Bishop}, D.~J., {Carr},
  E., {Cowan}, W.~D., {Hart}, M.~R., {Helmbrecht}, M., {Krulevitch}, P.~A.,
  {Muller}, R.~S., {Sadoulet}, B., {Solgaard}, O., and {Yu}, J.,
  ``{Micro-electro-mechanical systems spatial light modulator development},''
  in [{\em Society of Photo-Optical Instrumentation Engineers (SPIE) Conference
  Series}{\nolinebreak\hspace{0.1em}]},  {J.~D.~Gonglewski, M.~A.~Vorontsov, \&
  M.~T.~Gruneisen}, ed., {\em Society of Photo-Optical Instrumentation
  Engineers (SPIE) Conference Series} {\bf 4124},  26--31 (2000).

\bibitem{norton2009}
Norton, A., Evans, J.~W., Gavel, D., Dillon, D., Palmer, D., Macintosh, B.,
  Morzinski, K., and Cornelissen, S., ``Preliminary characterization of
  {B}oston {M}icromachines' 4096-actuator deformable mirror,'' {\em Proc.
  SPIE}~{\bf 7209},  72090I--72090I--7 (2009).

\bibitem{cfao2009}
CfAO, ``{CfAO} {Year} 9 {Progress} {Report}.''
  cfao.ucolick.org/pubs/Year\_9\_Annual\_Report.pdf (2009).

\bibitem{evans2006wavefront}
Evans, J.~W., Sommargren, G., Macintosh, B.~A., Severson, S., and Dillon, D.,
  ``Effect of wavefront error on $10^{-7}$ contrast measurements,'' {\em Optics
  Letters}~{\bf 31},  565--567 (2006).

\bibitem{evans2006flat}
Evans, J.~W., Macintosh, B., Poyneer, L., Morzinski, K., Severson, S., Dillon,
  D., Gavel, D., and Reza, L., ``Demonstrating sub-nm closed loop {MEMS}
  flattening,'' {\em Optics Express}~{\bf 14},  5558--5570 (2006).

\bibitem{evans2005spie}
Evans, J.~W., Morzinski, K., Reza, L., Severson, S., Poyneer, L., Macintosh,
  B., Dillon, D., Sommargren, G., Palmer, D., Gavel, D., and Olivier, S.,
  ``Extreme adaptive optics testbed: High contrast measurements with a {MEMS}
  deformable mirror,'' in [{\em Techniques and Instrumentation for Detection of
  Exoplanets II}{\nolinebreak\hspace{0.1em}]},  Coulter, D.~R., ed., {\em Proc.
  SPIE {\bf 5905}},  303--310 (2005).

\bibitem{evans2007}
{Evans}, J.~W., {Thomas}, S., {Dillon}, D., {Gavel}, D., {Phillion}, D., and
  {Macintosh}, B., ``{Amplitude variations on the ExAO testbed},'' in [{\em
  Society of Photo-Optical Instrumentation Engineers (SPIE) Conference
  Series}{\nolinebreak\hspace{0.1em}]},  {\em Society of Photo-Optical
  Instrumentation Engineers (SPIE) Conference Series} {\bf 6693} (Sept. 2007).

\bibitem{morzinski2006}
Morzinski, K.~M., Evans, J.~W., Severson, S., Macintosh, B., Dillon, D., Gavel,
  D., Max, C., and Palmer, D., ``Characterizing the potential of {MEMS}
  deformable mirrors for astronomical adaptive optics,'' in [{\em Advances in
  Adaptive Optics II}{\nolinebreak\hspace{0.1em}]},  Ellerbroek, B.~L. and
  Calia, D.~B., eds., {\em Proc. SPIE {\bf 6272}},  627221 (2006).

\bibitem{morzinski2009}
Morzinski, K., Macintosh, B., Gavel, D., and Dillon, D., ``Stroke saturation on
  a {MEMS} deformable mirror for woofer-tweeter adaptive optics,'' {\em Optics
  Express}~{\bf 17},  5829 (2009).

\bibitem{thomas2009}
Thomas, S., Evans, J.~W., Gavel, D., Dillon, D., and Macintosh, B., ``Amplitude
  variations on a mems-based extreme adaptive optics coronagraph testbed,''
  {\em Applied Optics}~{\bf 48},  4077 (2009).

\bibitem{severson2006}
Severson, S.~A., Bauman, B., Dillon, D., Evans, J., Gavel, D., Macintosh, B.,
  Morzinski, K., Palmer, D., and Poyneer, L., ``The {E}xtreme {A}daptive
  {O}ptics {T}estbed at {UCSC}: Current results and coronagraphic upgrade,'' in
  [{\em Advances in Adaptive Optics II}{\nolinebreak\hspace{0.1em}]},
  Ellerbroek, B.~L. and Calia, D.~B., eds., {\em Proc. SPIE {\bf 6272}},
  62722J (2006).

\bibitem{evans2009}
Evans, J.~W., Macintosh, B., Norton, A., Dillon, D., and Gavel, D., ``The
  effect of a small heat source on {PSF} stability for high-contrast imaging,''
  {\em Optics Express}~{\bf 17},  11652 (2009).

\bibitem{morzinski2008}
Morzinski, K.~M., Gavel, D., Norton, A., Dillon, D., and Reinig, M.,
  ``Characterizing {MEMS} deformable mirrors for open-loop operation:
  High-resolution measurements of thin-plate behavior,'' in [{\em {MEMS}
  Adaptive Optics II}{\nolinebreak\hspace{0.1em}]},  Olivier, S.~S., Bifano,
  T.~G., and Kubby, J.~A., eds., {\em Proc. SPIE {\bf 6888}},  68880S (2008).

\bibitem{bifano2000}
Bifano, T., Perreault, J., and Bierden, P., ``A micromachined deformable mirror
  for optical wavefront compensation,'' {\em Proc. SPIE}~{\bf 4124},  7--14
  (2000).

\bibitem{macintosh2006}
Macintosh, B., Graham, J., Palmer, D., Doyon, R., Gavel, D., Larkin, J.,
  Oppenheimer, B., Saddlemyer, L., Wallace, J.~K., Bauman, B., Evans, J.,
  Erikson, D., Morzinski, K., Phillion, D., Poyneer, L., Sivaramakrishnan, A.,
  Soummer, R., Thibault, S., and Veran, J.-P., ``The {G}emini {P}lanet
  {I}mager,'' in [{\em Advances in Adaptive Optics
  II}{\nolinebreak\hspace{0.1em}]},  Ellerbroek, B.~L. and Calia, D.~B., eds.,
  {\em Proc. SPIE {\bf 6272}},  62720L (2006).

\bibitem{morzinski2007}
Morzinski, K.~M., Harps{\o}e, K. B.~W., Gavel, D., and Ammons, S.~M., ``The
  open-loop control of {MEMS}: Modeling and experimental results,'' in [{\em
  {MEMS} Adaptive Optics}{\nolinebreak\hspace{0.1em}]},  Olivier, S., Bifano,
  T.~G., and Kubby, J.~A., eds., {\em Proc. SPIE {\bf 6467}},  64670G (2007).

\bibitem{stewart2007}
Stewart, J.~B., Diouf, A., Zhou, Y., and Bifano, T.~G., ``Open-loop control of
  a {MEMS} deformable mirror for large-amplitude wavefront control,'' {\em J
  Opt Soc Am A}~{\bf 24}(12),  3827--3833 (2007).

\bibitem{blain2000}
{Blain}, C., {Conan}, R., {Bradley}, C., and {Guyon}, O., ``{Open-loop control
  demonstration of micro-electro-mechanical-system {MEMS} deformable mirror},''
  {\em Optics Express}~{\bf 18},  5433 (Mar. 2010).

\bibitem{morzinski2010}
{Morzinski}, K., {Johnson}, L.~C., {Gavel}, D.~T., {Grigsby}, B., {Dillon}, D.,
  {Reinig}, M., and {Macintosh}, B.~A., ``{Performance of MEMS-based
  visible-light adaptive optics at Lick Observatory: closed- and open-loop
  control},'' {\em Proc. SPIE}~{\bf 7736},  773659 (2010).

\bibitem{shea2000}
Shea, H.~R., Gasparyan, A., White, C.~D., and Comizzoli, R.~B., ``Anodic
  oxidation and reliability of {MEMS} poly-silicon electrodes at high relative
  humidity and high voltages,'' in [{\em MEMS Reliability for Critical
  Applications}{\nolinebreak\hspace{0.1em}]},  Lawton, R.~A., ed., {\em Proc.
  SPIE {\bf 4180}},  117--122 (2000).

\bibitem{plass2003}
Plass, R.~A., Walraven, J.~A., Tanner, D.~M., and Sexton, F.~W., ``Anodic
  oxidation-induced delamination of the {SUMMiT} poly 0 to silicon nitride
  interface,'' {\em Proc. SPIE}~{\bf 4980},  81 (2003).

\bibitem{shea2004}
Shea, H.~R., Gasparyan, A., Chan, H.~B., Arney, S., Frahm, R.~E., Lopez, D.,
  Sungho, J., and McConnell, R.~P., ``Effects of electrical leakage currents on
  {MEMS} reliability and performance,'' {\em IEEE Transactions on Device and
  Materials Reliability} ,  198--207 (2004).

\bibitem{hon2007}
Hon, M., DelRio, F.~W., Carraro, C., and Maboudian, R., ``Effects of relative
  humidity and actuation voltage on {MEMS} reliability,'' {\em Solid-State
  Sensors, Actuators and Microsystems Conference} ,  367--370 (2007).

\end{thebibliography}

\end{document}